\documentclass[twocolumn,english,superscriptaddress,longbibliography,reprint,prb]{revtex4}
\pdfoutput=1
\usepackage[T1]{fontenc}
\usepackage{color}
\usepackage{bm}
\usepackage{amssymb}
\usepackage{graphicx}
\makeatletter
\usepackage{babel}

\makeatother

\begin{document}

\title{Phase diagram of hydrogen and a hydrogen-helium mixture at planetary conditions by  Quantum Monte Carlo simulations}

\author{Guglielmo Mazzola}
\email{gmazzola@phys.ethz.ch}

\affiliation{Theoretische Physik, ETH Zurich, 8093 Zurich, Switzerland}

\author{Ravit Helled}

\affiliation{Institute for Computational Science, 
University of Zurich, 8057 Zurich, Switzerland}

\author{Sandro Sorella}

\affiliation{International School for Advanced Studies (SISSA)  and INFM Democritos National Simulation Center, via Bonomea 265, 34136 , Italy} 

\begin{abstract} 
Understanding planetary interiors is directly linked to our ability of simulating exotic quantum mechanical systems such as hydrogen (H) and hydrogen-helium (H-He) mixtures at high pressures and temperatures\cite{guillot_interiors_2005}. 
Equations of State (EOSs) tables based on Density Functional Theory (DFT), are commonly used by planetary scientists, although this method  allows only for a qualitative description of the phase diagram\cite{knudson2015direct}, due to  an incomplete treatment of electronic interactions\cite{burke_perspective_2012}.
Here we report Quantum Monte Carlo (QMC) molecular dynamics simulations of pure H and H-He mixture. 
We calculate the first QMC EOS at 6000 K for an H-He mixture of a proto-solar composition, and show the crucial influence of He on the H metallization pressure.
Our results can be used to calibrate other EOS calculations and are very timely given the accurate determination of Jupiter's gravitational field from the NASA Juno mission and the effort to determine its structure\cite{Wahl2017}. 
\end{abstract}

\maketitle

Since a few decades the link between the uncertainty of the H EOS and the internal structure of 
Jupiter (and other gaseous planets) has been investigated and many efforts to model Jupiter's interior have 
been carried\cite{guillot_interiors_2005,nettelmann2012jupiter,Miguel2016,Militzer2016}. 
The computation of an EOS from first principles requires to solve a many-body quantum mechanical problem, a task which is beyond the currently available theoretical and computational capabilities. 
In practice, we must resort to several approximations. 
The first is to decouple the ionic and electronic problems and consider the ions as classical or quantum particles,  determining their motion by following the  Born-Oppenheimer potential energy surface.
The second approximation concerns the description of the electronic interaction and the exchange one, due to the Pauli exclusion principle.

The standard approach to EOS calculations relies on Density Functional Theory (DFT), which targets the tridimensional electronic density rather than the ($N_e$ electrons) many-body wave-function. Its success and simplicity have lead to a widespread application in materials science and to the development of several software packages which allow fast and reproducible calculations\cite{lejaeghere2016reproducibility}.
Although DFT is \emph{formally} exact, the explicit functional form to describe the exchange and correlation (XC) effects between electrons remains approximated\cite{burke_perspective_2012}.
Indeed, a systematic and efficient route to improve XC functional is still lacking. Therefore, in practical solid state calculations, benchmarks against experimental data, are often required to validate  the  XC functional used to   describe the system in a satisfactory manner.

\begin{figure}
\noindent \centering{}\includegraphics[width=1\columnwidth]{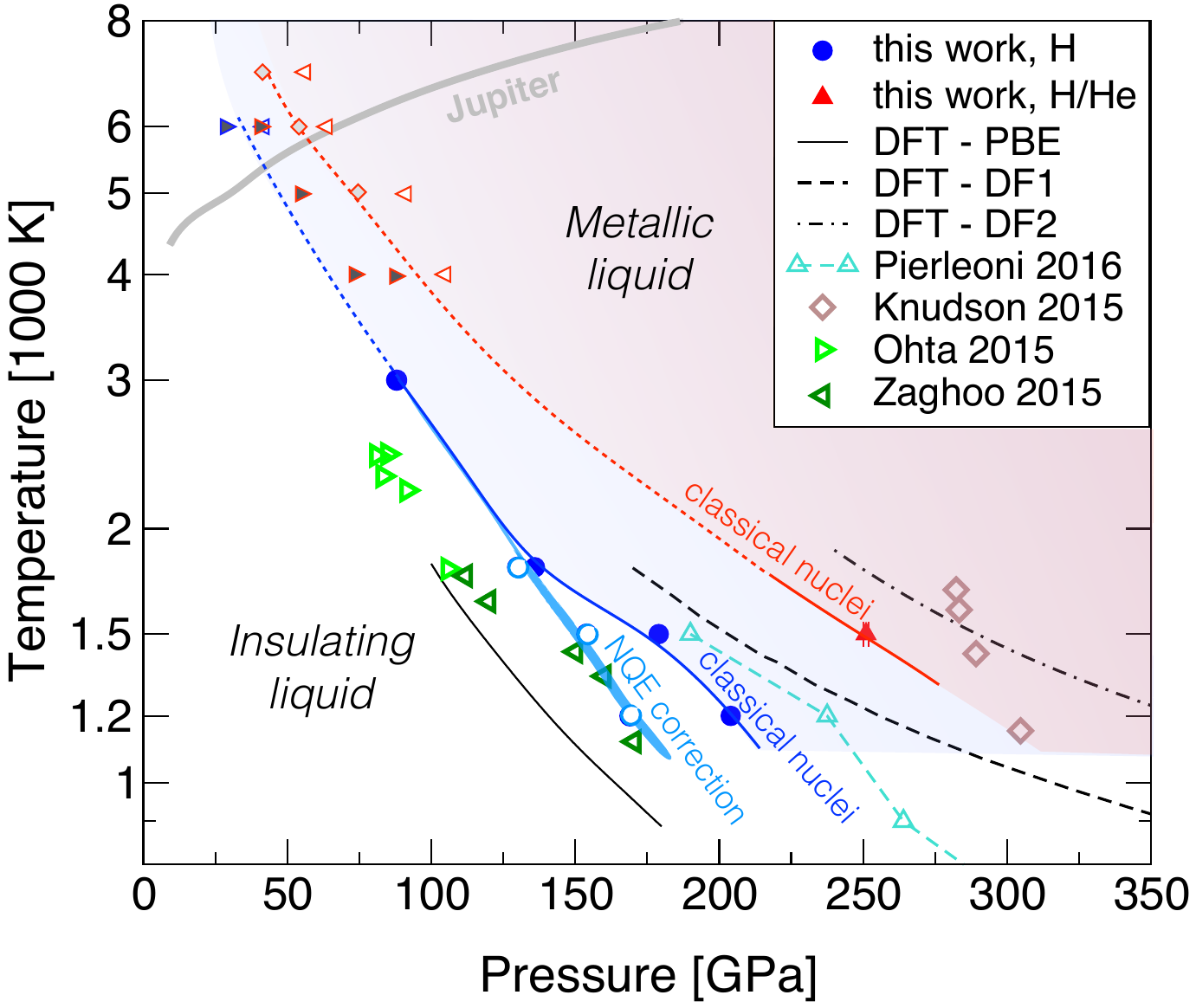}
\caption{\textbf{Phase diagram of dense hydrogen (H) and a hydrogen-helium (H-He) mixture.} We show the first-order liquid-liquid transition (LLT) between the insulating-molecular and the metallic-atomic fluid (shaded area), under the classical nuclei approximation.
Solid symbols refer to our QMC LLT for pure H (blue circles) and for the H-He mixture (red triangle).  
The light blue line sketches the position of the LLT if quantum nuclear effects (NQE) are included. In this case, the classical nuclei LLT (from our direct calculations, solid circles) is shifted according to Pierleoni et al.~\cite{pierleoni2016liquid} (empty circles, see text). 
At high temperatures, the empty (solid) left (right) triangle corresponds to simulations with a clear atomic (molecular) behaviour, while red diamonds represent an intermediate behaviour (see Methods). These points are used to constrain the phase boundaries (dashed lines) where a clear first-order transition is absent or difficult to identify.
Also shown is Jupiter's adiabat (grey line) as calculated by Miguel et al.~\cite{Miguel2016}.
We compare the LLT for classical nuclei pure H with recent QMC simulations of Pierleoni et. al..~\cite{pierleoni2016liquid} (cyan) and with DFT predictions using different XC functionals: PBE, vdW-DF1, and DF2 (with classical nuclei, taken from Knudson. et. al.~\cite{knudson2015direct}).  Other symbols refer to metallization experimental data. Shown are experiments with static compression~\cite{ohta2015phase,zaghoo2016evidence} (light and dark green triangles) and deuterium shockwave\cite{knudson2015direct} (brown diamonds). 
}
\label{Figure1} 
\end{figure}

Dense hydrogen systems, both in the low temperature solid and in the liquid phase remain a challenge to DFT simulations due to the interplay of strong correlation and non-covalent interactions between the atoms.  
DFT calculations with different functionals produce different results with the expected metallization pressure varying over a range of  100-200 GPa (Fig.~1)\cite{azadi_fate_2013,morales_nuclear_2013}.
Since experimental data are limited, it is not possible to identify \emph{a posteriori} the best functional for dense H.
Therefore the predictive power  expected by present ab-initio DFT simulations are somewhat limited.
Currently for Jupiter's interior, planetary modelers use H-He EOS that have been derived from DFT data\cite{militzer2013ab,becker2014ab}, using a specific -and somewhat arbitrary- choice  of the XC functional, the widely used Perdew-Burke-Ernzerhof (PBE)\cite{perdew_generalized_1996}.

Recently, QMC approaches emerged as competitive tools  to solve accurately  electronic problems\cite{Foulkes:2001p19717} thanks to the  
new generations of super-computers. 
Since QMC is a wave-function-based method (unlike DFT) the scheme to obtain consistently better results is simple and relies on the variational principle. 
Indeed, the accuracy  of the calculations  improves as the richness of the  many-body electronic wavefunction increases.
In our variational approach, a systematic way to improve the wavefunction is by enlarging  the localized atomic 
 basis set that defines our quantum state.
\begin{figure}
\noindent \centering{}\includegraphics[width=1\columnwidth]{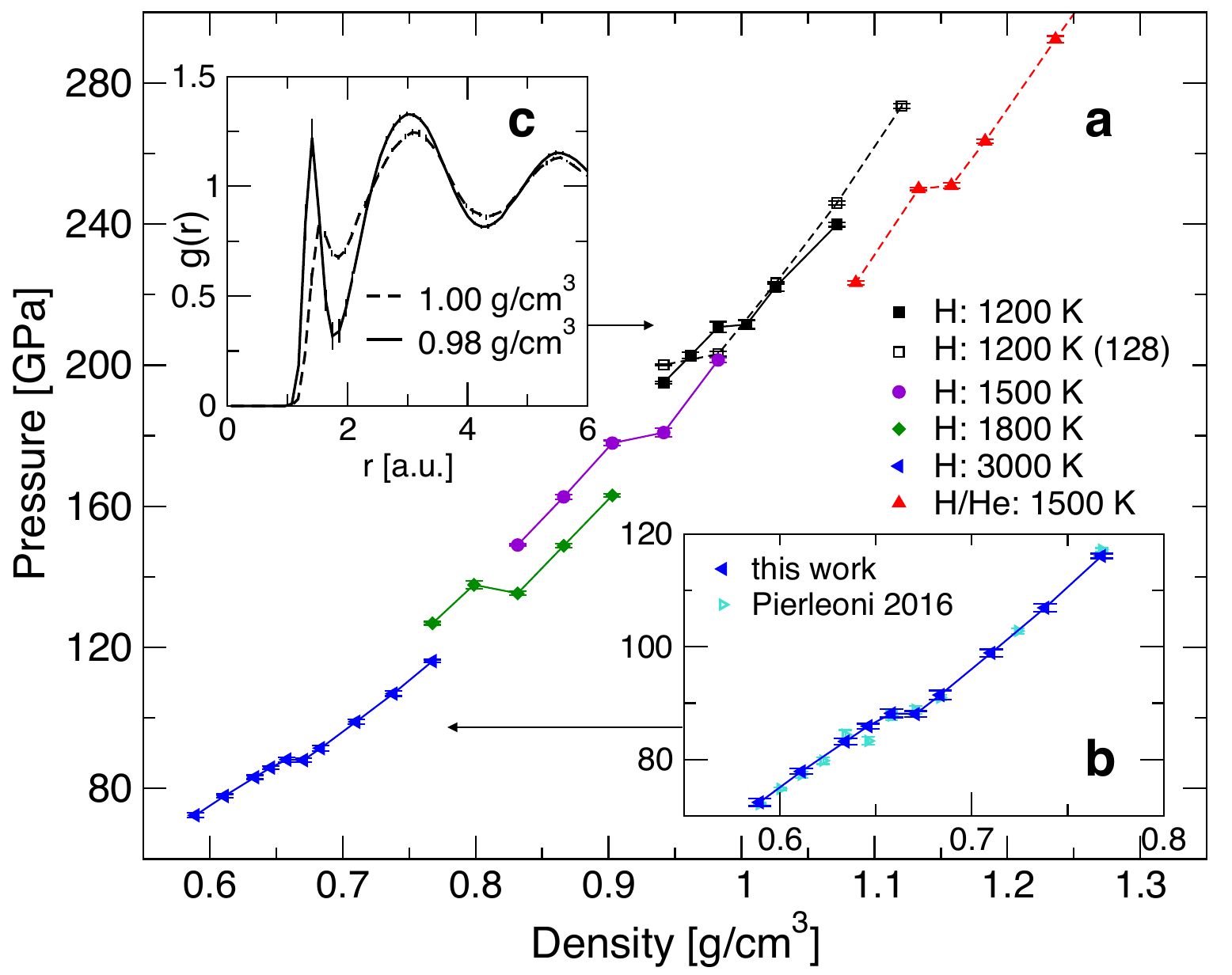}
\caption{\textbf{Equations of state across first-order transitions.} Pressure vs.~density  for pure H for four temperatures (1200, 1500, 1800 and 3000 K, 64 particles) and for a H-He mixture (at 1500 K, 128 particles). For H at 1200 K, we present results also for a 128 particles system.The first order transition is identified by a \emph{plateau} in the EOS. The discontinuity is more evident at lower temperatures but is still visible at 3000 K (panel {\bf a}). Panel {\bf b.} shows the EOS computed by Pierleoni et. al.~\cite{pierleoni2016liquid}.}
\label{Figure2} 
\end{figure}
The unprecedented availability of computational resources led to the development of QMC algorithms, that combine efficiently the simulations of electrons with ion dynamics\cite{attaccalite_stable_2008,Pierleoni:2004p28607,mazzola_unexpectedly_2014}.
Unlike the DFT method, which is well-established and widely used, the QMC technique is still relatively new and  
is used by a smaller community of developers with various  implementations and algorithms that are difficult to benchmark. 
However, the few QMC results for the H phase diagram, until now, have not agreed well.
In particular, while all QMC simulations agreed qualitatively on a larger dissociation and metallization pressure for pure dense liquid H, compared to PBE, the precise location was not well-determined due to different QMC implementations, variational wave-function, and finite size effects errors\cite{morales_evidence_2010,pierleoni2016liquid,mazzola_unexpectedly_2014,mazzola2015distinct,PhysRevLett.118.015703}.

We perform simulations with 64 and 128 H atoms for the H compound and with 118 H and 10 He atoms for the H-He mixture (see Methods). For the mixture we use  $x=n_{He}/n_{H}\approx 0.0926$ which is near the proto-solar value of $0.0969$\cite{RevModPhys.67.781} and slightly larger than Jupiter's value of $0.0785(18)$\cite{JGRE:JGRE847}. 
We first trace the liquid-liquid transition (LLT) for pure H at intermediate temperatures, between 1200K and 1800 K, using a 64 hydrogen atom system.
The first-order transition is characterized by a discontinuity in the EOS (see Fig.~2) and in the proton-proton radial pair distribution function $g(r)$ (Fig.~2c). It is found to occur at densities of $\sim 0.8-1$ g cm$^{-3}$ and pressures of $\sim 200$ GPa at $1200$ K, $\sim180$ GPa at $1500$ K, and $\sim 135$ GPa at $1800$ K.
The LLT seems to involve mostly a local-rearrangement of the liquid structure (see Fig.~2c). By comparing the $g(r)$s closest to the LLT, at densities around it from both sides, we notice the disappearance of the $H_2$ molecular peak at $1.4$ Bohr, and the appearance of a peak at $\approx 1.93$ Bohr, which corresponds to an equilibrium distance of the molecular $H_3^+$ ion\cite{norman2017critical} (see Supporting Figure S1). 
These features suggest the persistence of (possibly short-lived) molecular structures in the metallic fluid, which is not completely dissociated near the LLT.

Our QMC LLT  lies between the two recent 
experiments obtained using static compression by Silvera and 
coworkers\cite{dzyabura_evidence_2013,zaghoo2016evidence} and the dynamic compression measurements (with deuterium) by Knudson et al.\cite{knudson2015direct}, although it is much closer to the first reference.
Moreover, the systematic errors caused by the finite size and basis set can shift the LLT by $\sim$ 10 GPa, therefore, our results are compatible with the recent QMC prediction by Pierleoni et. al.\cite{pierleoni2016liquid}.
However, in order to better compare  our results with these low temperature experiments, also the quantum nature of the protons (here assumed as classical particles) needs to be considered.
Indeed, when we correct our results with these nuclear quantum effects (NQE), the agreement with static compression experiments improves significantly~\cite{fortov_phase_2007,dzyabura_evidence_2013,zaghoo2016evidence}(see Fig.1).
In this work we do not perform directly simulations beyond the classical nuclei approximation, using \emph{path integral} based methods as in Pierleoni et al.~\cite{pierleoni2016liquid}, where electronic QMC simulations with or without NQE are 
reported.
They show that NQE shifts the LLT to smaller pressures at most by $35$ GPa at $1200$ K and by $25$ GPa at $1500$ K. Here we simply apply these shifts to our LLT to derive the phase boundary in Fig.~1 and compare to experiments\cite{zaghoo2016evidence}. 
Notice that PBE underestimates the metallization pressure compared to QMC (Fig.~1), and the disagreement with experiments further increases if NQE are taken into account.

In this work, we correct the systematic errors that affected our previous results and led a much larger metallization pressure: the electronic size effects errors, not adequately removed in Mazzola et al.~\cite{mazzola_unexpectedly_2014,mazzola2015distinct} and a localized  basis set\cite{sorella2015geminal} that was too small (1Z) to describe 
 the metal and the insulator with the same accuracy\cite{PhysRevLett.118.015703}. 
 In addition, our previous studies used a less efficient optimization method,  indeed, the so called "linear method"\cite{umrigar_alleviation_2007} requires a careful generalization to the case of complex wavefunctions (see Methods for details).
 Nevertheless, our predicted LLT is affected by an uncertainty of $\sim 10$ GPa. We believe that, computing the H phase diagram with an accuracy of 1 GPa, is still beyond the present numerical capabilities, especially at low temperatures (see Methods and Supporting Figure S2).

\begin{figure}
\noindent \centering{}\includegraphics[width=1\columnwidth]{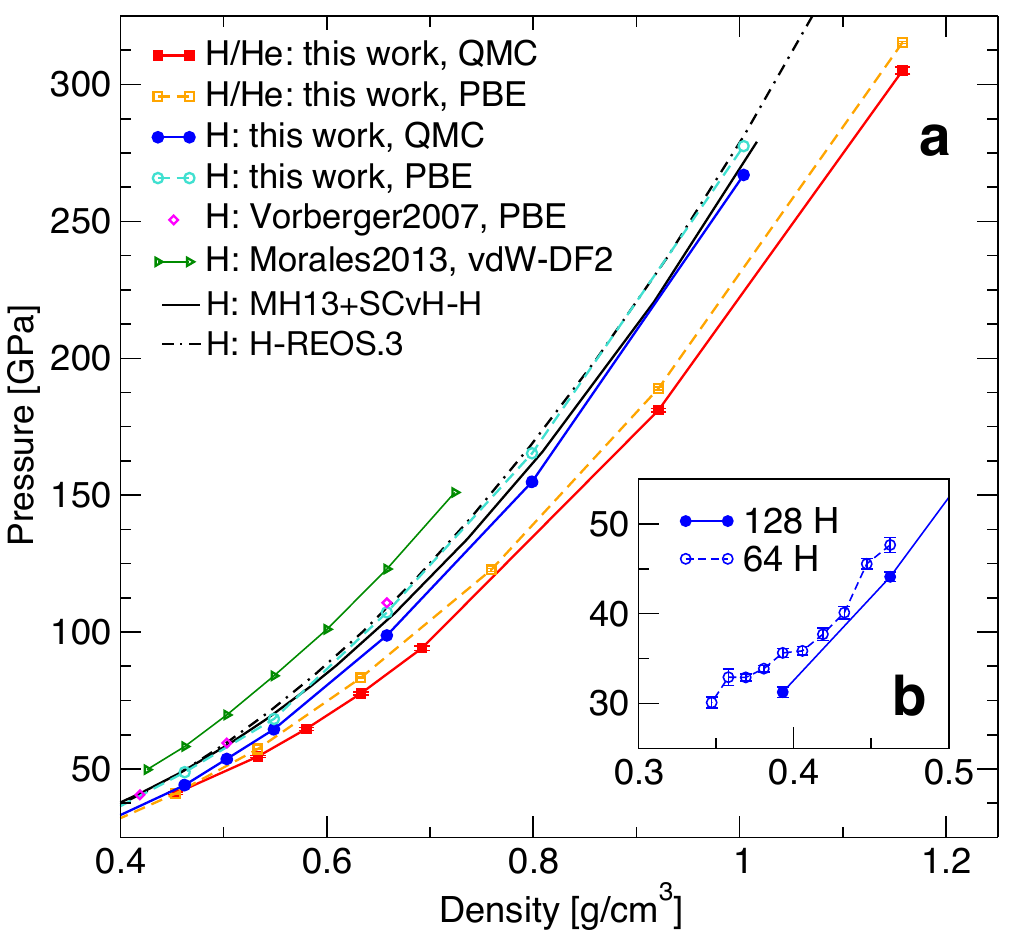}
\caption{\textbf{Equations of state at 6000 K.}  Panel {\bf a.} shows the EOS with QMC (solid lines) and DFT-PBE (dashed) for pure H (circles) and the H-He mixture (triangles) inferred using a supercell of 128 particles. For the pure H case we also report PBE calculations of Vorberger et. al.~\cite{PhysRevB.75.024206}, simulations with the vdW-DF2 functional,\cite{morales_nuclear_2013} and the commonly used EOSs for pure H, H-REOS.3\cite{becker2014ab} (dot-dashed black line) and MH-SCvH-H\cite{militzer2013ab} (continuos black line).
In the inset ({\bf b.}) we show additional simulations, using a 64 hydrogen system and a finer density mesh, to investigate the nature of the pure-H dissociation. 
Given the moderate slope and the statistical error bars ($\approx 0.5$ GPa) of the EOS, resolving any discontinuity is not possible at this stage.}
\label{Figure4} 
\end{figure}
\begin{figure}
\noindent \centering{}\includegraphics[width=1\columnwidth]{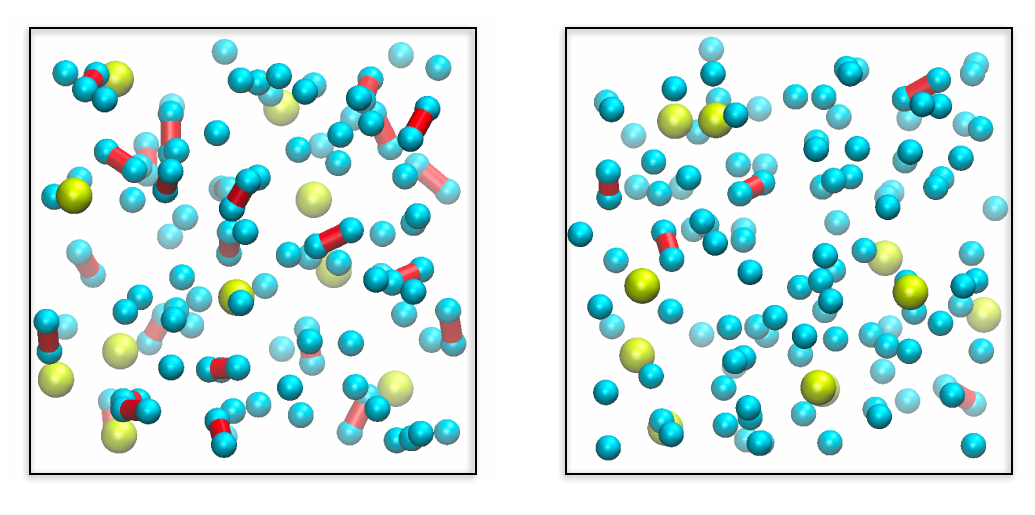} 
\caption{\textbf{Snapshots of an H-He mixture simulation.} Shown are two typical snapshots of our MD simulations.  
The cyan, yellow, and red colors represent the 108 H atoms, 10 He atoms and H-molecule bonds (H$_2$), respectively. For sake of visualization, an H$_2$ molecule is defined when two hydrogen atoms are closer than 1.6 Bohr. 
These two structures (projected on the \emph{x-y} plane), the first (second) represents a mainly molecular (atomic) phase, are computed at different iterations of the same MD simulation, at a temperature of 6000 K and density of $\sim$ 0.53 g/cm$^3$ (P = 54.3(3) GPa), i.e., near the LLT.
\label{snapshots} }
\end{figure}
After  benchmarking our technique for pure H, we next investigate an H-He mixture at $1500$ K.
We find that He, even in a small fraction $x \approx 0.093$, changes qualitatively the physics of the system.
In particular, its presence
stabilizes the hydrogen molecules (H$_2$), delaying the onset of metallization towards higher densities.
This effect is also observed in DFT-PBE simulations (cfn. Vorberger et. al.\cite{PhysRevB.75.024206}). However, our direct QMC simulations clearly identify the molecular dissociation in the H-He mixture at $\sim 250$ GPa with $1500$ K, and a density $\sim 1.1$ g cm$^{-3}$ (see Fig.~2a), resulting in a shift of $\approx 70$ GPa compared to the pure H system.

An important open question concerns the location of the H critical point, which is the end point of the first-order LLT. 
Above the critical point, in the $P-T$ phase diagram the dissociation occurs smoothly.
While recent EOS calculations suggest that Jupiter's adiabat lies above the critical point, implying the lack of first order phase transition, its possible occurrence has a direct consequence on the internal structure of gas giant planets. 
If the phase transition is of first order, it would suggest a density discontinuity within the planet's interior, and  the possibility of a non-adiabatic interior as well as for discontinues in the heavy elements distribution\cite{LC2012,Vazan2016}.

Also in this case a clear experimental consensus is still missing. McWilliams at. al.\cite{PhysRevLett.116.255501} do not find evidence for a first-order transition
   below 150 GPa, while
 Otha et. al.\cite{ohta2015phase} suggests instead the persistence of a first-order LLT well above 2000 K. Motivated by these studies, we perform an additional simulation at higher temperatures.
In the pure H case we are able to resolve a small discontinuity in the EOS and the $g(r)$ at 3000 K (see Fig.~2). 
Although a finer mesh of densities is required, as well as an extended finite size scaling in order to precisely resolve the existence of the \emph{plateau} in the EOS,
 the observed feature suggests the existence of a critical point above the previously expected temperature of 1500-2000 K\cite{Plorenzen,morales_evidence_2010} (with the notable exception of Norman et al.~\cite{norman2017critical} who predict a critical temperature of 4000 K from PBE simulations).

Finally, we calculate QMC and DFT-PBE EOSs at $6000$ K over a wide range of densities, spanning a pressure range between 30 (40) and 260 (300) GPa for pure H (H-He mixture). 
This isotherm is expected to cross Jupiter's adiabat around 60 GPa, i.e., at 0.6 Mbar.

We find that QMC, at a given density, predicts a pressure which is $\sim 5\%$ smaller than PBE, i.e., at a fixed pressure, QMC predicts a denser liquid compared to PBE. 
This difference becomes even larger when we compare against the vdW-DF2 results (see Fig.~3). 
Also shown in the figure is a comparison of our calculation with the two popular H-EOS for planetary interiors; the H-REOS.3\cite{becker2014ab} and MH-SCvH-H\cite{militzer2013ab}, both of them are based on PBE simulations. 
We show that H-REOS.3  is in perfect agreement with our DFT calculations, whereas the MH-SCvH-H EOS (extrapolating the data in the limiting case of pure H\cite{Miguel2016}) is closer to our QMC one. 
The disagreement between the two EOSs could be caused by the extrapolation\cite{Miguel2016}, and it seems that the disagreement between these two groups is linked to the calculated  entropies\cite{HM2016,Miguel2016}.
Either way, it is clear that QMC implies a denser EOS for H at Jupiter's conditions, which translates to an envelope that is poor in heavy elements. If this is indeed the case, it introduces new challenges in understanding Jupiter's current structure and origin\cite{Wahl2017, Miguel2016, HM2016}.

Regarding the nature of the phase transition at 6000 K, we find that for H our QMC simulation indicates that a crossover  is most likely to occur, as a clear EOS discontinuity is absent (see inset of Fig.~3).
This means that the critical temperature for pure H is between 3000 K and 6000 K. We can further constrain the location of the  LLT by performing simulations at different temperatures  and densities, identifying the largest(smallest) pressure at which a clear molecular peak persists(disappear).
For the H-He mixture, we directly perform simulations at temperatures between 4000 and 7000 K (see Supplementary Figures S3, S4),  relevant for planetary interiors (Fig.~1).
At 6000 K, H$_2$ dissociation in the H-He mixture occurs mainly between 42 and 64 GPa.  Moreover, at a density of $\sim$ 0.53 g cm$^{-3}$ and P $\approx$ 54 GPa, we observe the stability of  a mixed  phase  as the simulation quickly fluctuates between a pure atomic and a mainly molecular liquid (see Fig.~4). 
Therefore, our calculations show that the crossover from molecular to metallic hydrogen in Jupiter's conditions occurs at $\sim$ 0.4-0.6 Mbar. This provides further constraints for Jupiter structure models, as the transition pressure between the two envelopes cannot be used as a free parameter\cite{HG2013,Miguel2016}.  A transition pressure of that value implies a larger mass of heavy elements in Jupiter's deep interior\cite{Miguel2016}.

Our ab-initio simulations for an hydrogen-helium mixture obtained with QMC, an accurate electronic wave-function based method, opens an opportunity to better constrain the behavior of H and  H-He  in planetary conditions.  
We show that even a small concentration of He (at protosolar value of $\approx 9\%$) has an important impact on the metallization pressure of the liquid, as the dissociation is delayed by 70 GPa at low temperature (1500 K) and by 30 GPa at 6000 K.
While high pressure experiments employing mixtures are still lacking, the predicted phase boundary of the pure H compound is in good agreement with static compression experiments, at low temperatures.
New generation of QMC calculations, possibly tackling even larger systems, can reveal new information on the EOS of 
giant planets' deep interior, and address important questions such as the precise location of the critical point in the H-He phase diagram and the miscibility of He and other heavier materials in H.

\section{methods}

\subsection{General form of the variational wavefunction for QMC simulations}
We minimize the total energy expectation value of the
first-principles Hamiltonian, within the Born-Oppenheimer approximation,
by means of a correlated wave function, $J | SD \rangle$. This is made of a Slater
determinant (SD) defined in a localized basis
 and
a Jastrow term, containing two terms, a one-electron one
$J_1= \prod_{i} \exp{ \! \left [u_{1body}(\textbf{r}_i) \right] }$
and a two-electron correlation factor
 $J = \prod_{i<j} 
\exp{\!\left[u(\textbf{r}_i,\textbf{r}_j)\right]}$
explicitly dependent on the $N_e$ electronic positions, $\{\textbf{r}_i \}$,
and, parametrically, on the $N_e$
Hydrogen and Helium positions, $\textbf{R}_I$, $I=1,\cdots N_e$ and  valence charge 
$Z_I$ ($Z_I=1$ for Hydrogen and $Z_I=2$ for Helium).
The pseudopotential functions $u$ and $u_{1body}$ are written as:
\begin{eqnarray} \label{eq:jwf}
u(\textbf{ r},\textbf{ r}^\prime) &= &
  u_{ee} ( | \textbf{ r} - \textbf{ r}^\prime| ) 
  + \sum_{\mu>0,\nu>0} u_{\mu\nu}\, \chi_\mu(\textbf{r}) 
\chi_\nu(\textbf{r}^\prime)  \nonumber \\
 u_{1body}(\textbf{r})&=&u_{ei}(\textbf{r})+\sum_{\mu>0} u_{\mu,0} \chi_\mu(\textbf{r}),
\end{eqnarray}
where $u_{ee}$ and $u_{ei}$  are  simple functions:
\begin{eqnarray}
u_{ee}(r) &=& { r \over 2 ( 1 + b_{ee} r) } \nonumber \\
u_{ei}(\textbf{r}) &=& -\sum\limits_{I=1}^{N_{a}} Z_I (1 -\exp(-b_{ei}(Z_I) |{\bf r}-{\bf R}_I|))/b_{ei}(Z_I)
\end{eqnarray}
that allow us 
to satisfy the electron-electron and electron-ion cusp conditions, respectively, whereas 
$b_{ee}$ and $b_{ei}(Z_I)$ are few  variational parameters
 and $\textbf{u}$ is a symmetric matrix of finite dimension, determining the 
full variational freedom of the two functions $u$ and $u_{1body}$ 
 for large enough basis set.
Indeed the presence of the two simple functions $u_{ee}$ and $u_{ei}$, enormously accelerates the convergence to the complete basis set limit (CBS) as we will show in the following for the $u_{1body}$ within the DFT framework.
Notice that, the one-body Jastrow term $J_1$ is expanded in the same basis and its variational
freedom is determined by the first  column of the matrix $u$ for $\nu=0$.

In order to show that this approach is useful, not only within quantum Monte Carlo, 
we consider a system of 30 Hydrogen chain with periodic boundary conditions at fixed nearest neighbor distance of $2 a.u.$. This system was recently adopted for a benchmark study on the electron correlation problem (see Motta et. al.\cite{Motta_benchmark}).
 We use a systematically increasing basis set to describe the determinantal part, 
namely the $ccpVXT$ with $X=D,T, ...$.  The application of the one-body Jastrow 
$J_1$ to a Slater determinant  is equivalent to replace the corresponding 
molecular orbitals $\phi_i (\textbf{r})$ ($i=1,\cdots N_h/2$) by: 
$\phi_i (\textbf{r}) \to \exp( u_{1body} (\textbf{r})) \phi_i(\textbf{r})$ 
and this means that the standard basis of Gaussian type orbitals $ \phi_i (\textbf{r})$ 
is analogously modified, so that the electron-ion cusp conditions can be 
verified even with the simplest one body term $u_{1body} = u_{ei}$ described by only one 
variational parameter $b_{ei}(1)$.
We see in Supplementary Fig. ~S5 the remarkably fast convergence of our
approach, the CBS limit here represented by the standard Quantum Espresso value of 
the LDA functional for very large cutoff. 
We clearly see that the single $Z$ basis adopted in the previous works
is certainly unsatisfactory. However as soon as we increase the basis 
by using  the $ccpVDZ$ one  with the above modification, we find that,  
not only our modified $ccpVDZ$ basis is better and 
almost converged to the CBS limit (within $0.01eV/atom$) as compared to the standard $ccpVTZ$ one, but we can also remove the $p$ orbitals from this basis to obtain an essentially converged result for $b_{ei}(1)>\sim 1.91$. 
In the following we have adopted this scheme both for the Hydrogen ($b_{ei}(1)=1.91$) 
and Helium ($b_{ei}(2)=2.5$) atoms, by using three uncontracted $s$-wave gaussian orbitals for each atom 
$Z=1.962,0.44,0.122$ and $Z= 5.77,1.24,0.2976$ for the Hydrogen and Helium cases, respectively.
In order to be systematic, we have adopted the same basis set also for expanding the 
two functions $u$ and $u_{1body}$ corresponding to the Jastrow factor.

\subsection{Variational  optimization with complex wavefunctions }
Electronic finite size effects are removed using a $4\times4\times4$ k-points sampling of the Brillouin zone.
 The use of twisted boundary conditions implies that the variational 
wavefunction is generically complex in the determinantal part, because it is 
 described here  
by a Slater determinant with $N_e$ complex molecular orbitals $\phi_{i,\pm 1/2} ({\bf r}), i=1,N_e/2$  satisfying:
\begin{eqnarray}
 \phi_{i,\pm 1/2} \left({\bf r}+ (L,0,0)\right) &=&\exp( \pm i \theta_x )  \phi_i (\bf{r}) \nonumber \\
 \phi_{i,\pm 1/2}  \left({\bf r}+ (0,L,0)\right) &=&\exp( \pm i \theta_y )  \phi_i (\bf{r}) \nonumber \\
 \phi_i \left({\bf r}+ (0,0,L)\right) &=&\exp( \pm i \theta_z )  \phi_i (\bf{r}) \nonumber 
\end{eqnarray}
where $\theta_x,\theta_y,\theta_z$ are the twists used in the boundary 
conditions for a cubic box of volume $L^3$. 
The plus (minus) sign refers to the spin-up (spin-down)  molecular orbitals, so that the time reversal symmetry is satisfied. This requirement  has a considerable advantage in the implementation of the so called twisted averaged boundary conditions (TABC), as described in details in Dagrada et. al.\cite{refkturbo}.
The generalization of the so called ''linear method''\cite{umrigar_alleviation_2007} to a complex wavefunction has not been discussed in details so far.
Therefore we review here the basic steps and its generalization to this important case.
In the following we consider for simplicity that all parameters 
are real
because in case a complex parameter appear in the description of the molecular orbitals, we can consider its real and 
imaginary part as two independent real ones.
Instead the Jastrow factor is  assumed real and defined only in terms of real parameters.
In order to derive our main result it is convenient to express 
 the expectation value of the hamiltonian over 
the ansatz state 
$|\Psi_z \rangle$,  defined by a  {\em linear} correction to the 
wavefunction $|\Psi_{\alpha} \rangle$ with given $p$ 
variational parameters $\{ \alpha_k \}, k=1,\cdots,p$.
A small change of the variational parameters $\alpha_k \to \alpha_k + \delta \alpha_k$ produces on each electronic configuration $x$, where the electron positions and their spins are defined, 
a change that can be evaluated by  simple 
Taylor expansion:
\begin{equation}
\langle x |\Psi_{\mathbf{\alpha}+ \delta {\mathbf \alpha}} \rangle  = 
 \Psi_{\mathbf{\alpha}}(x) +\sum\limits_{k=1}^p \delta \alpha_k 
\partial_{\alpha_k} \Psi_{\mathbf{\alpha}}(x)   
\end{equation}
where, here and henceforth $ \Psi_{\mathbf{\alpha}}(x)  = \langle x | \Psi_{\mathbf{\alpha}} \rangle$. 
If we do not consider higher order terms we can therefore define the ''linear ansatz'' $\Psi_{\mathbf z}$ as:
\begin{equation}
\langle x |\Psi_{\mathbf z} \rangle = \sum\limits_{k=0}^p z_k O_k(x) \Psi_{\mathbf{\alpha}}(x)  
\end{equation}
where $\Psi_{\mathbf{\alpha}}(x) \ne 0$, and $O_k(x)$ is given by:
\begin{equation} \label{eq:defoi}
{\cal O}_k(x) = {\partial_{\alpha_k} \Psi_{\mathbf{\alpha}}(x) \over \Psi_{\mathbf{\alpha}} (x)}
\end{equation}
whereas ${\cal O}_0(x)=1$. 
Here the vector components $z_k$, defining $\Psi_{\mathbf z}$, 
are real  
as a consequence of the assumption that the variational parameters are all real.
The Hamiltonian expectation value over $\Psi_z$ is therefore 
given by:
\begin{equation} \label{hexp}
{\langle \Psi_z | H | \Psi_z \rangle \over 
\langle \Psi_z | \Psi_z \rangle } = {\sum\limits_{k,k^\prime} z_k z_{k^\prime} H_{k,k^\prime} \over \sum\limits_{k,k^\prime} S_{k,k^\prime} z_k z_{k^\prime}} 
\end{equation}
and can be evaluated by computing  
stochastically  two matrices $H_{k,k^\prime}$ and $S_{k,k^\prime}$, whose leading dimension is $p+1$: 
\begin{eqnarray}
&& S_{k,k^\prime} \approx \frac{1}{N} \sum_{i=1}^N {\cal O}_k(x_i)^{*}
{\cal O}_{k^\prime}(x_i), \\
&& H_{k,k^\prime} \approx \frac{1}{N} \sum_{i=1}^N {\cal O}_{k}(x_i)^{*}
\frac{\langle x_i|{\cal H} {\cal O}_{k^\prime}|\Psi_{\alpha}\rangle}
{\langle x_i|\Psi_{\alpha}\rangle}.
\end{eqnarray}
The restriction to real valued $z_k$, which is unavoidable due to the presence 
of the Jastrow,  implies that 
we have to take 
the real part of both 
matrices $S$ and $H$, (this operation does not change the expectation value in Eq.~\ref{hexp}, because $H$ is hermitian) 
as a restriction of the more general complex valued 
$\{ z_k \}$ ansatz, implying that:
\begin{eqnarray}
H_{k,k^\prime} &\to & {\bar H}_{k,k^\prime} = {\cal R} (H_{k,k^\prime}) \nonumber \\ 
S_{k,k^\prime} & \to & {\bar H}_{k,k^\prime} ={\cal R} (S_{k,k^\prime}).
\end{eqnarray} 
Within this restriction,  
we are naturally lead  to the following {\rm real} generalized 
eigenvalue problem:
\begin{equation}\label{eq:gendiag}
\sum_{k^\prime=0}^p {\bar H}_{k,k^\prime} z_{k^\prime} = E \sum_{k^\prime=0}^p {\bar S}_{k,k^\prime} z_{k^\prime},
\end{equation}
that can be solved by standard linear algebra packages. Once the right 
eigenvector 
corresponding to the minimum energy eigenvalue $E$  is evaluated,
 $z_i$ define the new variational parameters at the next iteration 
by $ \delta \alpha_k = s {z_k \over z_0}$, where $s$ is a scaling factor
that is conveniently determined empirically, in order to have a more stable optimization scheme, as the straightforward choice $s=1$, determined by the 
assumption that non linear terms in the Taylor expansion can be neglected, 
 is often unstable. 
Here we have used $s=0.35 {z_0 \over z_0 + \sum_{k>0} z_k {\bar O}_k }$, where ${\bar O}_k= 1/N {\cal R}(\sum_{i} O_k( x_i))$. 
Notice that for infinite statistics the matrix $H_{k,k^\prime}$ is symmetric but it is not convenient to symmetrize it.
Indeed, only without symmetrizing it  the so called ''strong zero variance property'' can be satisfied because if $\Psi_z$ is an eigenstate of 
$H$ with eigenvalue $E$, the generalized eigenvalue equation is exactly fullfilled even within a simulation with a finite number of samples $N\ge p+1$. 
Indeed if $\Psi_{\mathbf z}$ is an eigenstate then:
\begin{widetext}
\begin{eqnarray}
\sum\limits_{k^\prime} {\bar H}_{k,k^\prime} z_k &=& {\cal R}\left[
 \frac{1}{N} \sum_{i=1}^N {\cal O}_{k}(x_i)^{*}
\frac{\langle x_i|{\cal H} \sum\limits_{k^\prime} z_{k^\prime} {\cal O}_{k^\prime}|\Psi_{\alpha}\rangle}
{\langle x_i|\Psi_{\alpha}\rangle} \right]= 
{\cal R}\left[
 \frac{1}{N} \sum_{i=1}^N {\cal O}_{k}(x_i)^{*}
\frac{\langle x_i|{\cal H} |\Psi_{\mathbf z}\rangle}
{\langle x_i|\Psi_{\alpha}\rangle} \right] \nonumber \\
&=& E
{\cal R}\left[
 \frac{1}{N} \sum_{i=1}^N {\cal O}_{k}(x_i)^{*}
\sum\limits_{k^\prime} z_k {\cal O}_{k^\prime}(x_i)  \right] 
= E \sum\limits_{k^\prime} {\bar S}_{k,k^\prime} z_{k^\prime}
\end{eqnarray}
\end{widetext}
which proves the mentioned statement, whenever we consider at least $N=p+1$ 
independent configurations, because otherwise the matrices ${\bar H}$ and 
${\bar S}$ are rank-deficient and the generalized eigenvalue equation is not well defined.

In the Hydrogen case we have verified that the optimization of the determinantal part provides an irrelevant improvement of the energy, i.e. much smaller than 1mH/atom and therefore, in this case, we have optimized only the Jastrow, using ten iterations of the linear method for  each step  of molecular dynamics, that safely allows us to remain with the electronic wavefunction
 very close to  the Born-Oppheneimer energy surface, namely at the minimum possible energy given by our variational ansatz, with optimal values for the matrix 
$\textbf{u}$, and reoptimized coefficients 
 $b_{ei}(1)$, and $b_{ee}$.
On the contrary in the mixture we have found a more significant role of the optimization.
Though in this case the improvement in the energy provided by the determinantal part remains small (of the order of 1mH/atom) we have optimized also this part  toghether with  the Jastrow (same parameters as before and $b_{ei}(2)$).
In this case, in order to deal with a reasonably small number of parameters, we restrict the optimization of the determinant within  local atomic corrections to  the DFT initial determinant, which, we have verified, takes into account most of the energy improvement.
Moreover, since the optimization is a bit slower when the determinant is optimized, we have 
used in this case twenty iterations of the linear method 
for each step of Molecular dynamics.

\subsection{QMC Molecular dynamics setup}
We have used the recently introduced accelerated molecular dynamics\cite{PhysRevLett.118.015703} to sample the classical canonical distribution in the NVT ensemble.
In this dynamics a crucial role is played by the parameter $\Delta_0$ that is used to scale 
the covariance matrix $S$ used for the acceleration in order to represent as close as 
possible the Hessian matrix. This is helpful to obtain a very small error in the time step
$\Delta$. For any choice of $\Delta_0$ the limit $\Delta \to 0$ gives the exact sampling 
at finite temperature, but for the appropriate choice we can use much larger time steps.
In Supplementary Fig.~S6 we show the internal energy and the pressure 
as a function of the time step for a system of $N_h=64$ Hydrogens in the atomic phase.
It is clear that for $\Delta$ small enough all the choices of $\Delta_0$ provide internal
energies consistent within 0.01 eV/atom and pressures within 1 GPa. However 
for $\Delta_0=$500 Ha$^{-1}$ the dependence on the time step is almost optimal (flat) and we have 
therefore adopted this value for all the  simulations in Hydrogen.
Moreover we have used  a small time 
step $\Delta =$ 0.0002 Ha, so that we do not need any extrapolation in the time step as, by judging from this plot, the error should be in any event negligible. 

Notice that in this dynamics the time step has the same dimension of the energy, due to the multiplication of the forces for the 
 acceleration matrix, that has dimension 
$-2$ in the energy.

\subsection{DFT Molecular dynamics setup}

DFT simulations are performed using Perdew, Burke and Ernzerhof (PBE)\cite{perdew_generalized_1996} functional on a 128 particle system, we use a 3x3x3 Monkhorst-Pack k-point sampling mesh, with pseudopotentials of the projector-augmented wave type (for both hydrogen and helium) and a plane wave basis set cutoff of 60 Ry (increased up to 80 Ry in the case of mixtures and for larger densities).
The ionic sampling is driven by a very simple first-order Langevin thermostat $R' = R + \Delta f + \eta$, with $\Delta$ in the range 0.2-0.4 Ry$^{-1}$ (notice that the friction coefficient is put to 1 Bohr$^{-2}$ as a different value only rescales the time step), depending on the temperature and density parameters.
A comparison with the Andersen thermostat has been made for selected densities and temperatures.
The electronic entropy is  taken into account approximatively by using the \emph{smearing} of the Fermi occupation number distribution. 
We notice that the approximate inclusion of the electronic entropy with the smearing gives a negligible difference compared with simulations without smearing, suggesting that the ground-state approximations holds very well even at the largest temperature here considered.
Overall, we notice that our DFT simulations are in perfect agreement with previous studies using PBE\cite{PhysRevB.75.024206,Plorenzen,becker2014ab}.



\section{Additional figures}

\begin{figure}[]
\noindent \centering{}\includegraphics[width=1\columnwidth]{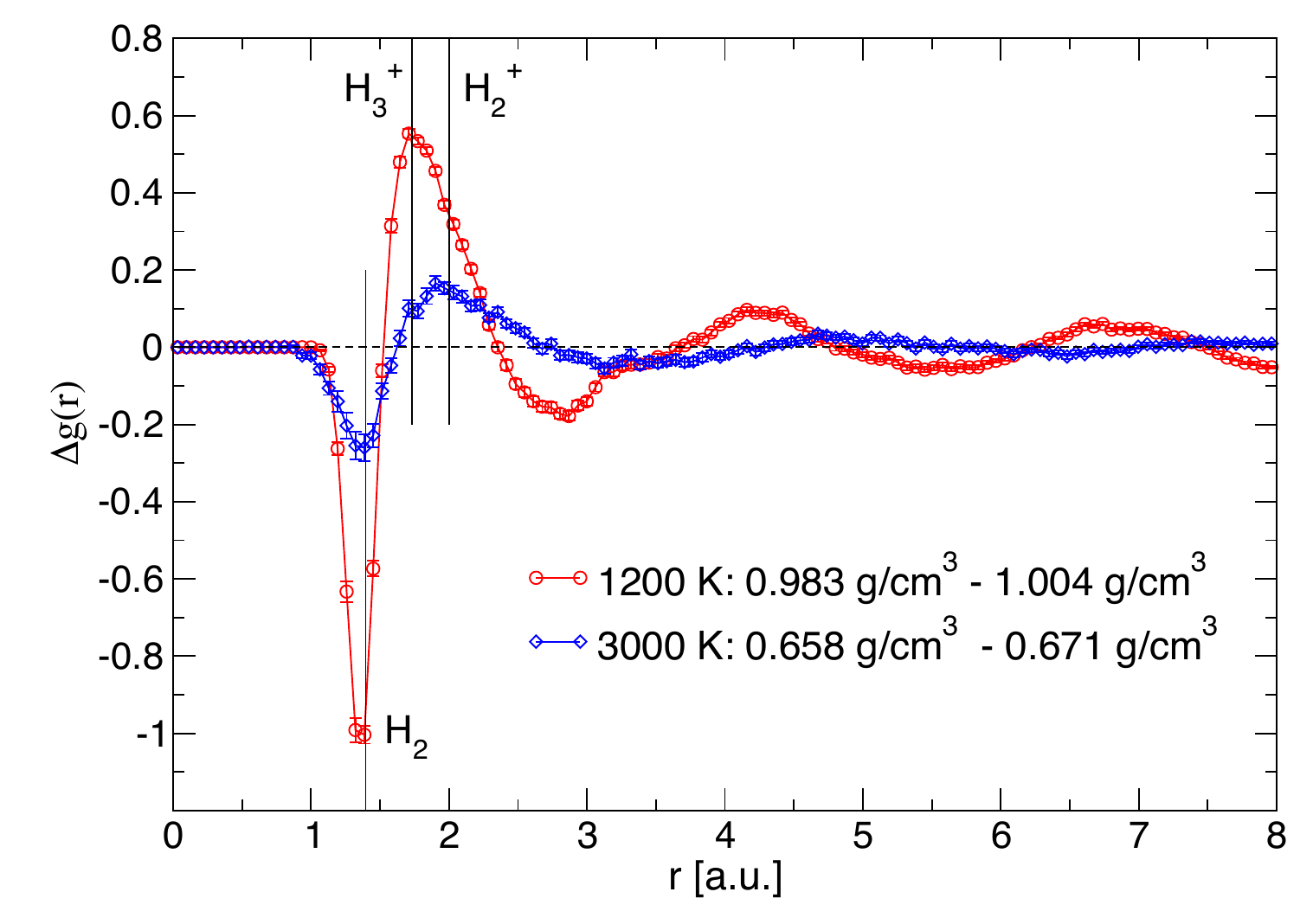}
\caption{{\bf Supplementary Figure} S1: Change of the radial pair distribution function $g(r)$ at the first-order transition. We denote with $\Delta g(r)$ the  difference between the two $g(r)$'s corresponding to  the densities (indicated in the figure legend) closest to the phase transition, before and after it\cite{norman2017critical}. The change in the liquid structure is more evident at low temperature (1200 K). The rearrangements of the liquid are mainly short range, as they involve the dissociation of the hydrogen molecules. The vertical lines mark the distances between the atoms for the hydrogen molecule ($H_2$) and for the molecular $H_3^+$ (1.74 Bohr) and $H_2^+$ (2.00 Bohr) ions.}
\label{timesteperror_atomic}
\end{figure}

\begin{figure}[]
\noindent \centering{}\includegraphics[width=1\columnwidth]{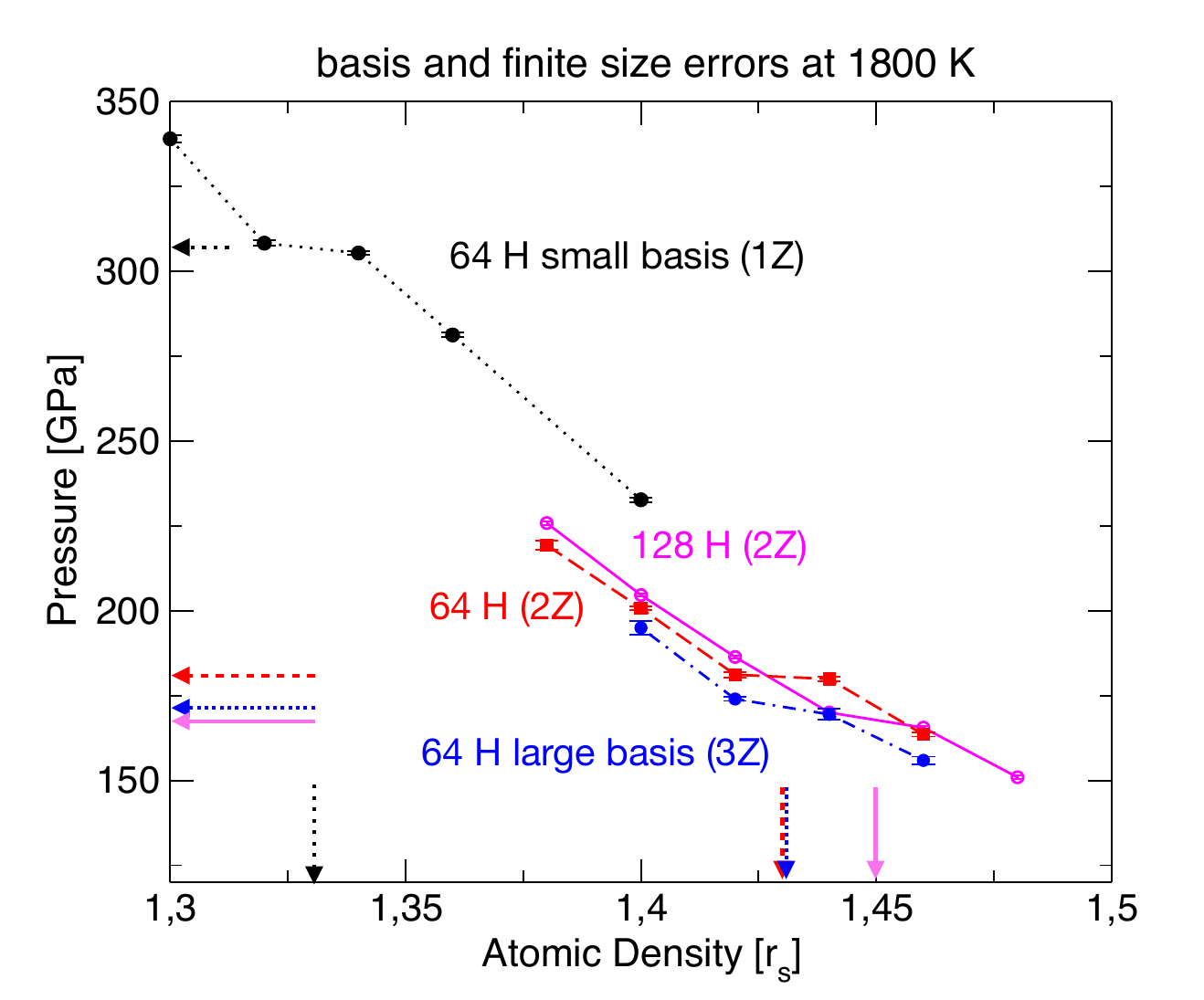}
\caption{{\bf Supplementary Figure} S2: \textbf{Basis set and finite size errors.} EOS calculated with different basis sets at 1800 K for a 64 hydrogen system with a time step larger than the one used to produce the final results. We label the particle density using the  Wigner-Seitz radius parameter $r_s$, defined as $V/N=4/3\pi r_s^3$, where $N$ is the number of \emph{atoms} and  $V$ is the volume of the simulation cell, expressed in atomic units ($1$ a.u. $\approx  0.053$ nm). The location of the LLT is sensitive to the basis set. The basis set used in this work (2Z) produces converged results within the reasonable accuracy of $ <$ 10 GPa. with respect to the much larger 3Z basis. The small 1Z basis, used in previuos works, produce an error larger than 100 GPa instead. We also notice a residual dependency of the LLT density as a function of the system size. Altough electronic finite size effects are treated with the k-point sampling of the Brillouin zone, a small deviation between the 64 and the 128 system is still present (for the 2Z basis set). This error cannot be completely removed with the technique used in Pierleoni et. al.\cite{pierleoni2016liquid} as different sizes produce also different average atomic geometries.}
\label{basiserror} 
\end{figure}

\begin{figure}[]
\noindent \centering{}\includegraphics[width=1\columnwidth]{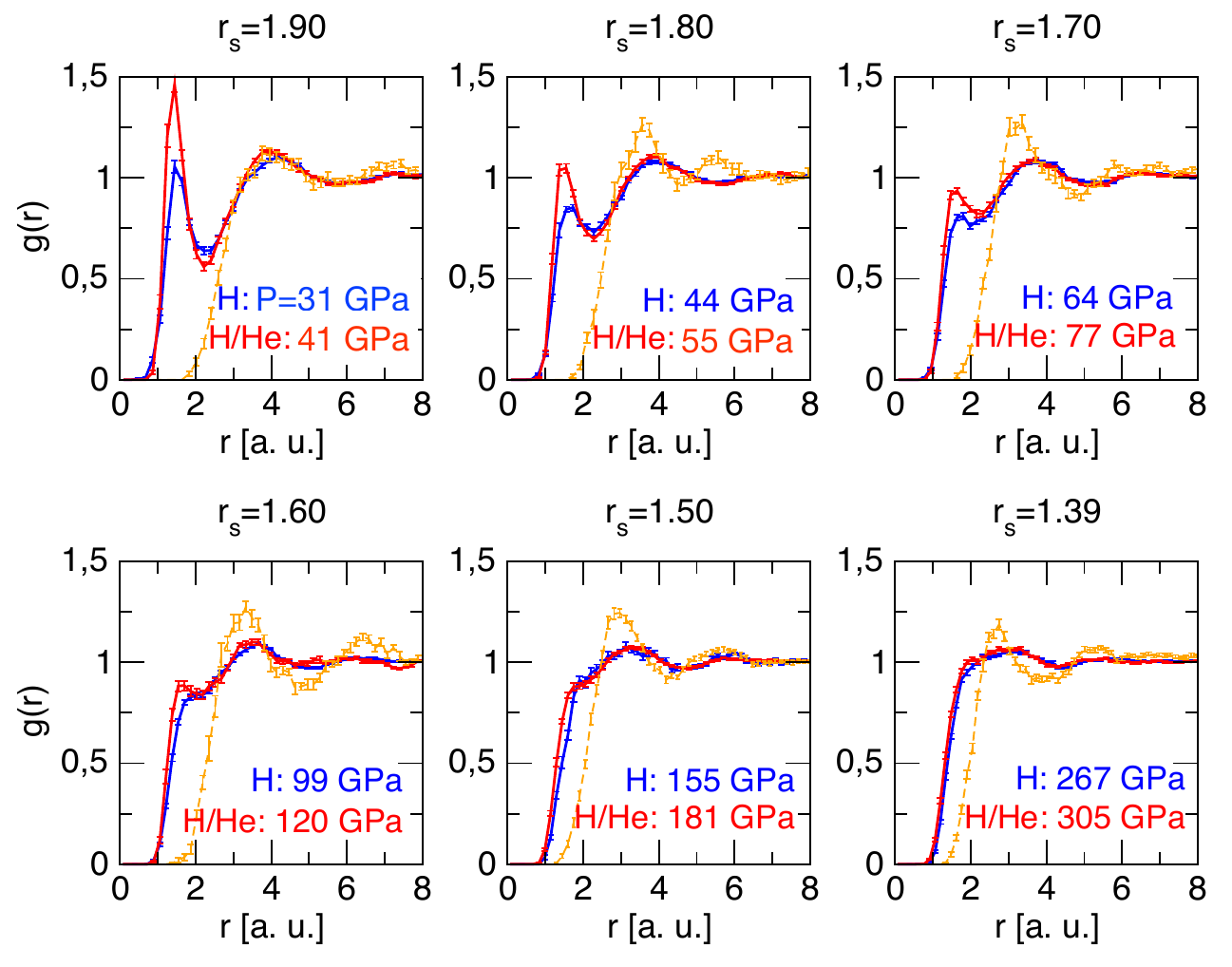}
\caption{{\bf Supplementary Figure} S3: Radial pair distribution functions $g(r)$ for QMC simulations for H and H-He for different densities (128 particle). We label the particle density using the  Wigner-Seitz radius parameter $r_s$, defined as $V/N=4/3\pi r_s^3$, where $N$ is the number of \emph{atoms} and  $V$ is the volume of the simulation cell, expressed in atomic units ($1$ a.u. $\approx  0.053$ nm).  Blue lines refer to the H-H $g(r)$ in pure hydrogen simulations, while red lines refer to H-H $g(r)$ in mixture simulations, while orange dashed line to the H-He distribution. At fixed particle density, the presence of He, stabilizes the hydrogen molecules (cfn. the peak at $\sim $ 1.4 a.u.). We also indicate the pressures of the H (blue) and H-He (red) corresponding to those densities.}
\label{mistura6000}
\end{figure}

\begin{figure}[]
\noindent \centering{}\includegraphics[width=1\columnwidth]{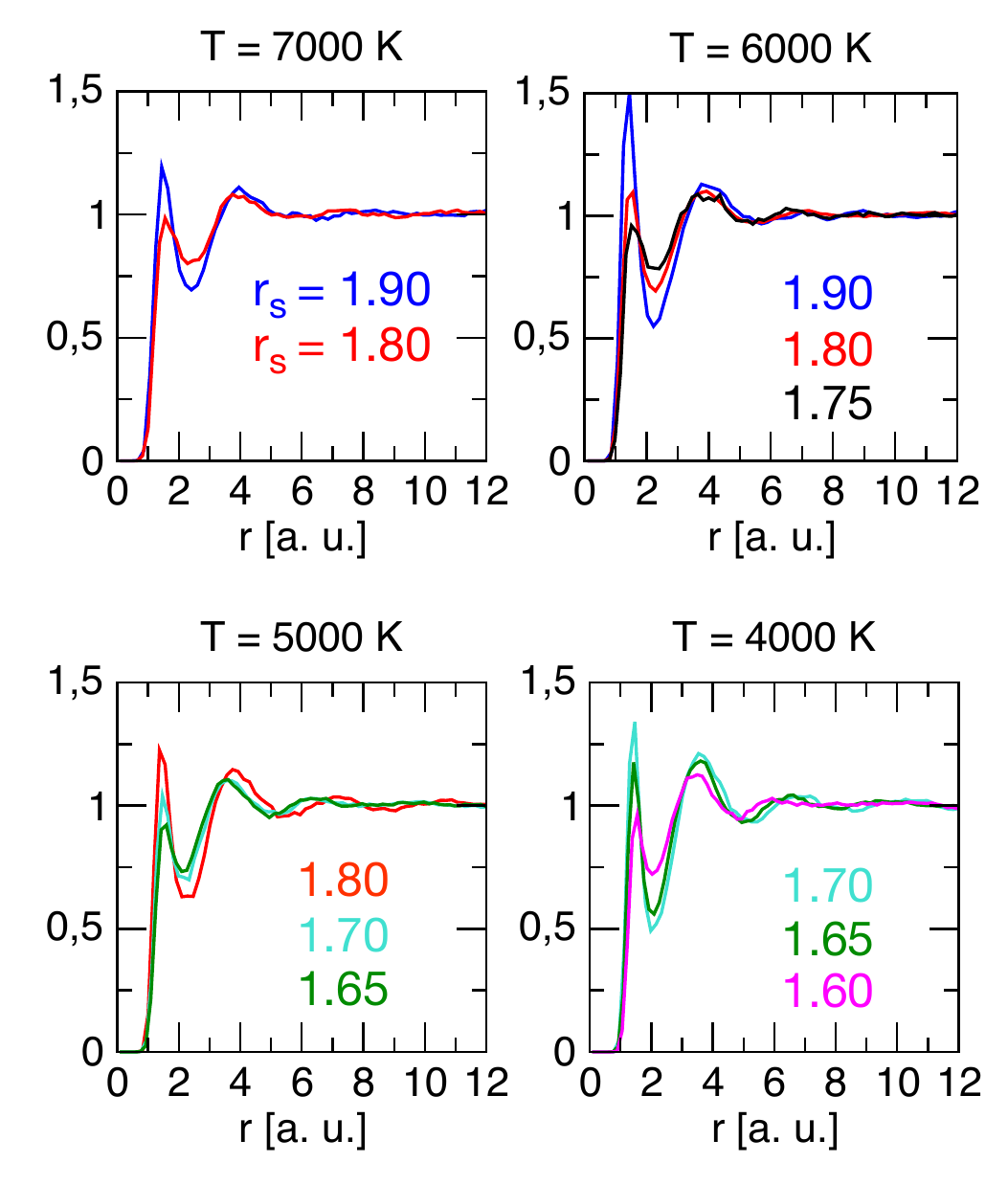}
\caption{{\bf Supplementary Figure} S4: Radial pair distribution functions $g(r)$ for H-He QMC simulations for different temperatures. We plot $g(r)$ at different particle densities (defined using the Wigner-Seitz radius defined above) at each temperature in order to constrain the metallization/dissociation crossover. }
\label{misturacrossover}
\end{figure}

\begin{figure}[]
\noindent \centering{}\includegraphics[width=1\columnwidth]{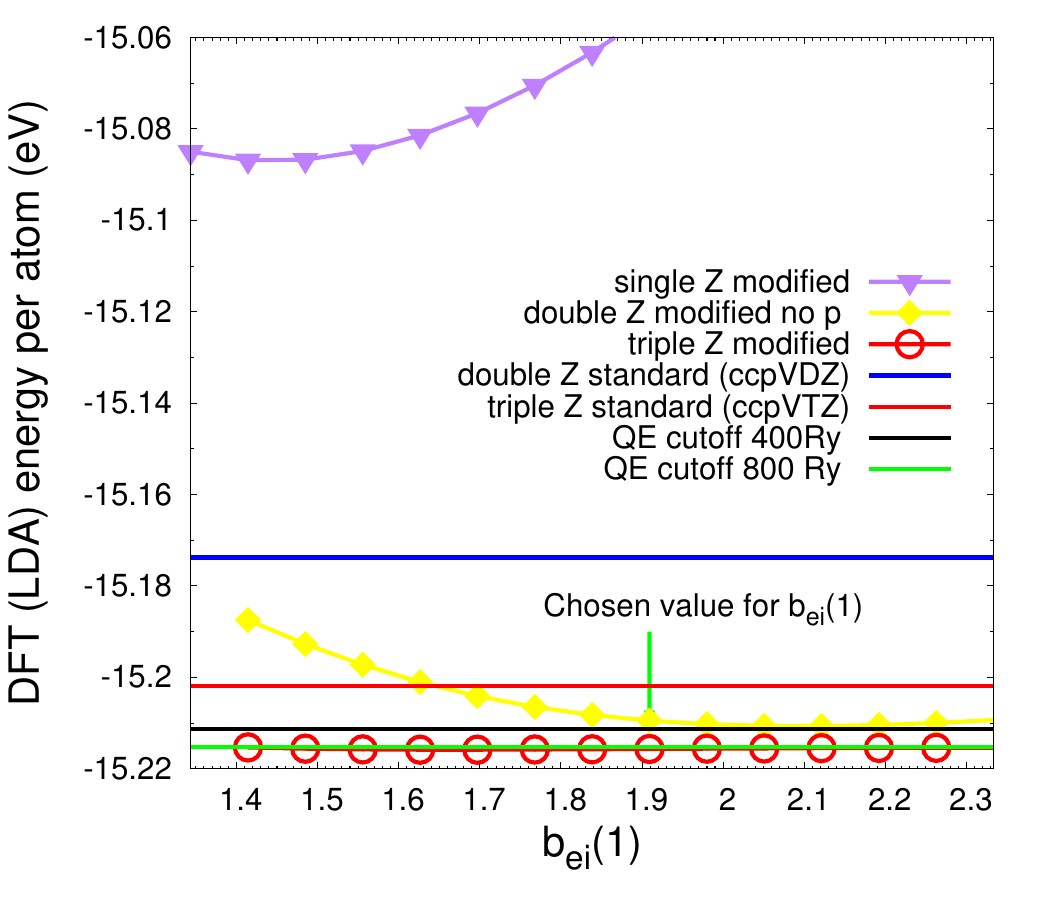}
\caption{{\bf Supplementary Figure} S5: Basis set convergence in DFT within the standard LDA functional.The single Z basis refers to the one adopted in the previous work\cite{PhysRevLett.118.015703}, where the accelerated molecular dynamics was introduced. $2Z$ and $3Z$ basis sets refer instead to the more conventional ''correlation consistent polarized valence $x-$zeta'' (ccpVxZ , with $x=2,3$, respectively) sequence\cite{ccpVXZ}. Our modification allows us  to satisfy the electron-ion cusp condition, by introducing only one variational parameter $b_{ei}$ that is weakly dependent on the electronic configuration, as it represents a property of the atomic core. The plane wave basis set depends only on the kinetic energy cutoff, taken large enough for converged results, and is obtained by a standard electronic structure package\cite{QE-2009}} 
\label{basiserrordft}
\end{figure}


\begin{figure}[]
\noindent \centering{}\includegraphics[width=1\columnwidth]{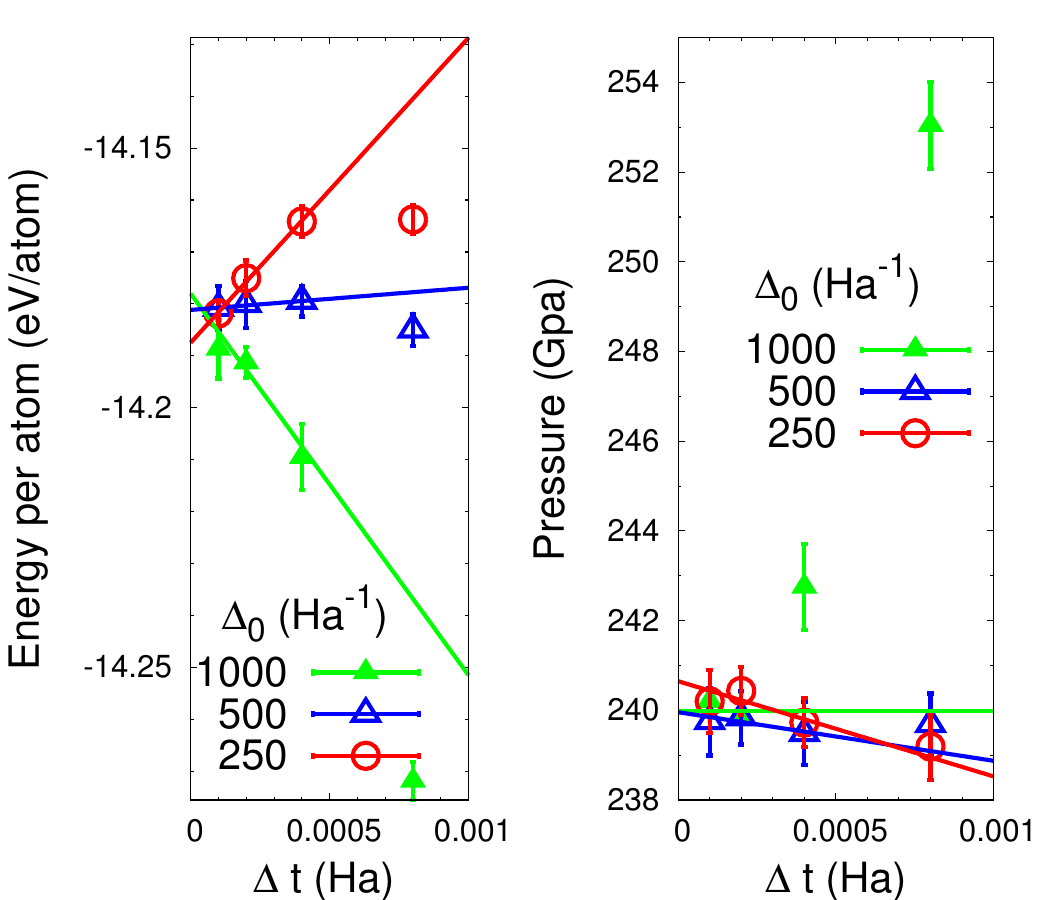}
\caption{{\bf Supplementary Figure} S6: Time step error in the atomic phase of Hydrogen at $r_s=1.36$ and $T=1200K$ for $64$ atoms. $\Delta_0$ here is a parameter that is used to optimize the time step error $\Delta t$ of the molecular dynamics\cite{PhysRevLett.118.015703}. For small enough $\Delta t$ all results should be unbiased. We have adopted therefore $\Delta_0=500$ ($\Delta_0=1000$)  and $\Delta t< 0.0002Ha$ in all  the production runs  used  in this work for determining the Hydrogen (Helium) phase diagram.}
\label{timesteperror_atomic}
\end{figure}

\clearpage

 {\bf Acknowledgements}

 G.M. was supported by the European Research Council through ERC Advanced Grant SIMCOFE, and the Swiss National Science Foundation through the National Competence Centers in Research MARVEL and QSIT. RH acknowledges support from the Swiss National Science Foundation (SNSF) project S-76104-01. Computational resources were provided by CINECA in Bologna, AICS-Riken project n.hp170079,  and by the Swiss National Supercomputing Centre CSCS.


\begin{thebibliography}{10}
\expandafter\ifx\csname url\endcsname\relax
  \def\url#1{\texttt{#1}}\fi
\expandafter\ifx\csname urlprefix\endcsname\relax\def\urlprefix{URL }\fi
\providecommand{\bibinfo}[2]{#2}
\providecommand{\eprint}[2][]{\url{#2}}

\bibitem{guillot_interiors_2005}
\bibinfo{author}{Guillot, T.}
\newblock \bibinfo{title}{{THE} {INTERIORS} {OF} {GIANT} {PLANETS:} models and
  outstanding questions}.
\newblock \emph{\bibinfo{journal}{Annual Review of Earth and Planetary
  Sciences}} \textbf{\bibinfo{volume}{33}}, \bibinfo{pages}{493--530}
  (\bibinfo{year}{2005}).

\bibitem{knudson2015direct}
\bibinfo{author}{Knudson, M.} \emph{et~al.}
\newblock \bibinfo{title}{Direct observation of an abrupt insulator-to-metal
  transition in dense liquid deuterium}.
\newblock \emph{\bibinfo{journal}{Science}} \textbf{\bibinfo{volume}{348}},
  \bibinfo{pages}{1455--1460} (\bibinfo{year}{2015}).

\bibitem{burke_perspective_2012}
\bibinfo{author}{Burke, K.}
\newblock \bibinfo{title}{Perspective on density functional theory}.
\newblock \emph{\bibinfo{journal}{The Journal of Chemical Physics}}
  \textbf{\bibinfo{volume}{136}}, \bibinfo{pages}{150901}
  (\bibinfo{year}{2012}).

\bibitem{Wahl2017}
\bibinfo{author}{{Wahl}, S. e.~a.}
\newblock \bibinfo{title}{{Comparing Jupiter interior structure models to Juno
  gravity measurements and the role of a dilute core}}.
\newblock \emph{\bibinfo{journal}{Geophys. Res. Lett.,}}
  \textbf{\bibinfo{volume}{44}}, \bibinfo{pages}{4649} (\bibinfo{year}{2017}).
\newblock \eprint{1602.05143}.

\bibitem{nettelmann2012jupiter}
\bibinfo{author}{Nettelmann, N.}, \bibinfo{author}{Becker, A.},
  \bibinfo{author}{Holst, B.} \& \bibinfo{author}{Redmer, R.}
\newblock \bibinfo{title}{Jupiter models with improved ab initio hydrogen
  equation of state (h-reos. 2)}.
\newblock \emph{\bibinfo{journal}{The Astrophysical Journal}}
  \textbf{\bibinfo{volume}{750}}, \bibinfo{pages}{52} (\bibinfo{year}{2012}).

\bibitem{Miguel2016}
\bibinfo{author}{{Miguel}, Y.}, \bibinfo{author}{{Guillot}, T.} \&
  \bibinfo{author}{{Fayon}, L.}
\newblock \bibinfo{title}{{Jupiter internal structure: the effect of different
  equations of state}}.
\newblock \emph{\bibinfo{journal}{A\&A}} \textbf{\bibinfo{volume}{596}},
  \bibinfo{pages}{A114} (\bibinfo{year}{2016}).
\newblock \eprint{1609.05460}.

\bibitem{Militzer2016}
\bibinfo{author}{{Militzer}, B.}, \bibinfo{author}{{Soubiran}, F.},
  \bibinfo{author}{{Wahl}, S.~M.} \& \bibinfo{author}{{Hubbard}, W.}
\newblock \bibinfo{title}{{Understanding Jupiter's interior}}.
\newblock \emph{\bibinfo{journal}{Journal of Geophysical Research (Planets)}}
  \textbf{\bibinfo{volume}{121}}, \bibinfo{pages}{1552--1572}
  (\bibinfo{year}{2016}).
\newblock \eprint{1608.02685}.

\bibitem{lejaeghere2016reproducibility}
\bibinfo{author}{Lejaeghere, K.} \emph{et~al.}
\newblock \bibinfo{title}{Reproducibility in density functional theory
  calculations of solids}.
\newblock \emph{\bibinfo{journal}{Science}} \textbf{\bibinfo{volume}{351}},
  \bibinfo{pages}{aad3000} (\bibinfo{year}{2016}).

\bibitem{pierleoni2016liquid}
\bibinfo{author}{Pierleoni, C.}, \bibinfo{author}{Morales, M.~A.},
  \bibinfo{author}{Rillo, G.}, \bibinfo{author}{Holzmann, M.} \&
  \bibinfo{author}{Ceperley, D.~M.}
\newblock \bibinfo{title}{Liquid--liquid phase transition in hydrogen by
  coupled electron--ion monte carlo simulations}.
\newblock \emph{\bibinfo{journal}{Proceedings of the National Academy of
  Sciences}} \textbf{\bibinfo{volume}{113}}, \bibinfo{pages}{4953--4957}
  (\bibinfo{year}{2016}).

\bibitem{ohta2015phase}
\bibinfo{author}{Ohta, K.} \emph{et~al.}
\newblock \bibinfo{title}{Phase boundary of hot dense fluid hydrogen.}
\newblock \emph{\bibinfo{journal}{Scientific reports}}
  \textbf{\bibinfo{volume}{5}}, \bibinfo{pages}{16560--16560}
  (\bibinfo{year}{2015}).

\bibitem{zaghoo2016evidence}
\bibinfo{author}{Zaghoo, M.}, \bibinfo{author}{Salamat, A.} \&
  \bibinfo{author}{Silvera, I.~F.}
\newblock \bibinfo{title}{Evidence of a first-order phase transition to
  metallic hydrogen}.
\newblock \emph{\bibinfo{journal}{Physical Review B}}
  \textbf{\bibinfo{volume}{93}}, \bibinfo{pages}{155128}
  (\bibinfo{year}{2016}).

\bibitem{azadi_fate_2013}
\bibinfo{author}{Azadi, S.} \& \bibinfo{author}{Foulkes, W. M.~C.}
\newblock \bibinfo{title}{Fate of density functional theory in the study of
  high-pressure solid hydrogen}.
\newblock \emph{\bibinfo{journal}{Physical Review B}}
  \textbf{\bibinfo{volume}{88}}, \bibinfo{pages}{014115}
  (\bibinfo{year}{2013}).

\bibitem{morales_nuclear_2013}
\bibinfo{author}{Morales, M.~A.}, \bibinfo{author}{{McMahon}, J.~M.},
  \bibinfo{author}{Pierleoni, C.} \& \bibinfo{author}{Ceperley, D.~M.}
\newblock \bibinfo{title}{Nuclear quantum effects and nonlocal
  exchange-correlation functionals applied to liquid hydrogen at high
  pressure}.
\newblock \emph{\bibinfo{journal}{Physical Review Letters}}
  \textbf{\bibinfo{volume}{110}}, \bibinfo{pages}{065702}
  (\bibinfo{year}{2013}).

\bibitem{militzer2013ab}
\bibinfo{author}{Militzer, B.} \& \bibinfo{author}{Hubbard, W.~B.}
\newblock \bibinfo{title}{Ab initio equation of state for hydrogen-helium
  mixtures with recalibration of the giant-planet mass-radius relation}.
\newblock \emph{\bibinfo{journal}{The Astrophysical Journal}}
  \textbf{\bibinfo{volume}{774}}, \bibinfo{pages}{148} (\bibinfo{year}{2013}).

\bibitem{becker2014ab}
\bibinfo{author}{Becker, A.} \emph{et~al.}
\newblock \bibinfo{title}{Ab initio equations of state for hydrogen (h-reos. 3)
  and helium (he-reos. 3) and their implications for the interior of brown
  dwarfs}.
\newblock \emph{\bibinfo{journal}{The Astrophysical Journal Supplement Series}}
  \textbf{\bibinfo{volume}{215}}, \bibinfo{pages}{21} (\bibinfo{year}{2014}).

\bibitem{perdew_generalized_1996}
\bibinfo{author}{Perdew, J.~P.}, \bibinfo{author}{Burke, K.} \&
  \bibinfo{author}{Ernzerhof, M.}
\newblock \bibinfo{title}{Generalized gradient approximation made simple}.
\newblock \emph{\bibinfo{journal}{Physical Review Letters}}
  \textbf{\bibinfo{volume}{77}}, \bibinfo{pages}{3865--3868}
  (\bibinfo{year}{1996}).

\bibitem{Foulkes:2001p19717}
\bibinfo{author}{Foulkes, W.}, \bibinfo{author}{Mitas, L.},
  \bibinfo{author}{Needs, R.} \& \bibinfo{author}{Rajagopal, G.}
\newblock \bibinfo{title}{{Quantum Monte Carlo simulations of solids}}.
\newblock \emph{\bibinfo{journal}{Reviews Of Modern Physics}}
  \textbf{\bibinfo{volume}{73}}, \bibinfo{pages}{33--83}
  (\bibinfo{year}{2001}).

\bibitem{attaccalite_stable_2008}
\bibinfo{author}{Attaccalite, C.} \& \bibinfo{author}{Sorella, S.}
\newblock \bibinfo{title}{Stable liquid hydrogen at high pressure by a novel ab
  initio molecular-dynamics calculation}.
\newblock \emph{\bibinfo{journal}{Physical Review Letters}}
  \textbf{\bibinfo{volume}{100}}, \bibinfo{pages}{114501}
  (\bibinfo{year}{2008}).

\bibitem{Pierleoni:2004p28607}
\bibinfo{author}{Pierleoni, C.}, \bibinfo{author}{Ceperley, D.~M.} \&
  \bibinfo{author}{Holzmann, M.}
\newblock \bibinfo{title}{{Coupled electron-ion Monte Carlo calculations of
  dense metallic hydrogen}}.
\newblock \emph{\bibinfo{journal}{Physical Review Letters}}
  \textbf{\bibinfo{volume}{93}}, \bibinfo{pages}{146402}
  (\bibinfo{year}{2004}).

\bibitem{mazzola_unexpectedly_2014}
\bibinfo{author}{Mazzola, G.}, \bibinfo{author}{Yunoki, S.} \&
  \bibinfo{author}{Sorella, S.}
\newblock \bibinfo{title}{Unexpectedly high pressure for molecular dissociation
  in liquid hydrogen by electronic simulation}.
\newblock \emph{\bibinfo{journal}{Nature Communications}}
  \textbf{\bibinfo{volume}{5}}, \bibinfo{pages}{3487} (\bibinfo{year}{2014}).

\bibitem{morales_evidence_2010}
\bibinfo{author}{Morales, M.~A.}, \bibinfo{author}{Pierleoni, C.},
  \bibinfo{author}{Schwegler, E.} \& \bibinfo{author}{Ceperley, D.~M.}
\newblock \bibinfo{title}{Evidence for a first-order liquid-liquid transition
  in high-pressure hydrogen from ab initio simulations}.
\newblock \emph{\bibinfo{journal}{Proceedings of the National Academy of
  Sciences}} \textbf{\bibinfo{volume}{107}}, \bibinfo{pages}{12799--12803}
  (\bibinfo{year}{2010}).

\bibitem{mazzola2015distinct}
\bibinfo{author}{Mazzola, G.} \& \bibinfo{author}{Sorella, S.}
\newblock \bibinfo{title}{Distinct metallization and atomization transitions in
  dense liquid hydrogen}.
\newblock \emph{\bibinfo{journal}{Physical Review Letters}}
  \textbf{\bibinfo{volume}{114}}, \bibinfo{pages}{105701}
  (\bibinfo{year}{2015}).

\bibitem{PhysRevLett.118.015703}
\bibinfo{author}{Mazzola, G.} \& \bibinfo{author}{Sorella, S.}
\newblock \bibinfo{title}{Accelerating ab initio molecular dynamics and probing
  the weak dispersive forces in dense liquid hydrogen}.
\newblock \emph{\bibinfo{journal}{Phys. Rev. Lett.}}
  \textbf{\bibinfo{volume}{118}}, \bibinfo{pages}{015703}
  (\bibinfo{year}{2017}).

\bibitem{RevModPhys.67.781}
\bibinfo{author}{Bahcall, J.~N.}, \bibinfo{author}{Pinsonneault, M.~H.} \&
  \bibinfo{author}{Wasserburg, G.~J.}
\newblock \bibinfo{title}{Solar models with helium and heavy-element
  diffusion}.
\newblock \emph{\bibinfo{journal}{Rev. Mod. Phys.}}
  \textbf{\bibinfo{volume}{67}}, \bibinfo{pages}{781--808}
  (\bibinfo{year}{1995}).

\bibitem{JGRE:JGRE847}
\bibinfo{author}{von Zahn, U.}, \bibinfo{author}{Hunten, D.~M.} \&
  \bibinfo{author}{Lehmacher, G.}
\newblock \bibinfo{title}{Helium in jupiter's atmosphere: Results from the
  galileo probe helium interferometer experiment}.
\newblock \emph{\bibinfo{journal}{Journal of Geophysical Research: Planets}}
  \textbf{\bibinfo{volume}{103}}, \bibinfo{pages}{22815--22829}
  (\bibinfo{year}{1998}).

\bibitem{norman2017critical}
\bibinfo{author}{Norman, G.} \& \bibinfo{author}{Saitov, I.}
\newblock \bibinfo{title}{Critical point and mechanism of the fluid--fluid
  phase transition in warm dense hydrogen}.
\newblock In \emph{\bibinfo{booktitle}{Doklady Physics}},
  vol.~\bibinfo{volume}{62}, \bibinfo{pages}{294--298}
  (\bibinfo{organization}{Springer}, \bibinfo{year}{2017}).

\bibitem{dzyabura_evidence_2013}
\bibinfo{author}{Dzyabura, V.}, \bibinfo{author}{Zaghoo, M.} \&
  \bibinfo{author}{Silvera, I.~F.}
\newblock \bibinfo{title}{Evidence of a liquidâliquid phase transition in hot
  dense hydrogen}.
\newblock \emph{\bibinfo{journal}{Proceedings of the National Academy of
  Sciences}} \textbf{\bibinfo{volume}{110}}, \bibinfo{pages}{8040--8044}
  (\bibinfo{year}{2013}).
\newblock \bibinfo{note}{{PMID:} 23630287}.

\bibitem{fortov_phase_2007}
\bibinfo{author}{Fortov, V.~E.} \emph{et~al.}
\newblock \bibinfo{title}{Phase transition in a strongly nonideal deuterium
  plasma generated by quasi-isentropical compression at megabar pressures}.
\newblock \emph{\bibinfo{journal}{Physical Review Letters}}
  \textbf{\bibinfo{volume}{99}}, \bibinfo{pages}{185001}
  (\bibinfo{year}{2007}).

\bibitem{sorella2015geminal}
\bibinfo{author}{Sorella, S.}, \bibinfo{author}{Devaux, N.},
  \bibinfo{author}{Dagrada, M.}, \bibinfo{author}{Mazzola, G.} \&
  \bibinfo{author}{Casula, M.}
\newblock \bibinfo{title}{Geminal embedding scheme for optimal atomic basis set
  construction in correlated calculations}.
\newblock \emph{\bibinfo{journal}{The Journal of chemical physics}}
  \textbf{\bibinfo{volume}{143}}, \bibinfo{pages}{244112}
  (\bibinfo{year}{2015}).

\bibitem{umrigar_alleviation_2007}
\bibinfo{author}{Umrigar, C.~J.}, \bibinfo{author}{Toulouse, J.},
  \bibinfo{author}{Filippi, C.}, \bibinfo{author}{Sorella, S.} \&
  \bibinfo{author}{Hennig, R.~G.}
\newblock \bibinfo{title}{Alleviation of the fermion-sign problem by
  optimization of many-body wave functions}.
\newblock \emph{\bibinfo{journal}{Physical Review Letters}}
  \textbf{\bibinfo{volume}{98}}, \bibinfo{pages}{110201}
  (\bibinfo{year}{2007}).

\bibitem{PhysRevB.75.024206}
\bibinfo{author}{Vorberger, J.}, \bibinfo{author}{Tamblyn, I.},
  \bibinfo{author}{Militzer, B.} \& \bibinfo{author}{Bonev, S.~A.}
\newblock \bibinfo{title}{Hydrogen-helium mixtures in the interiors of giant
  planets}.
\newblock \emph{\bibinfo{journal}{Phys. Rev. B}} \textbf{\bibinfo{volume}{75}},
  \bibinfo{pages}{024206} (\bibinfo{year}{2007}).

\bibitem{LC2012}
\bibinfo{author}{{Leconte}, J.} \& \bibinfo{author}{{Chabrier}, G.}
\newblock \bibinfo{title}{{A new vision of giant planet interiors: Impact of
  double diffusive convection}}.
\newblock \emph{\bibinfo{journal}{Astronomy \& Astrophysics}}
  \textbf{\bibinfo{volume}{540}}, \bibinfo{pages}{A20} (\bibinfo{year}{2012}).
\newblock \eprint{1201.4483}.

\bibitem{Vazan2016}
\bibinfo{author}{{Vazan}, A.}, \bibinfo{author}{{Helled}, R.},
  \bibinfo{author}{{Podolak}, M.} \& \bibinfo{author}{{Kovetz}, A.}
\newblock \bibinfo{title}{{The Evolution and Internal Structure of Jupiter and
  Saturn with Compositional Gradients}}.
\newblock \emph{\bibinfo{journal}{The Astrophysical Journal}}
  \textbf{\bibinfo{volume}{829}}, \bibinfo{pages}{118} (\bibinfo{year}{2016}).
\newblock \eprint{1606.01558}.

\bibitem{PhysRevLett.116.255501}
R.~S. McWilliams, D.~A. Dalton, M.~F. Mahmood, and A.~F. Goncharov.
Optical Properties of Fluid Hydrogen at the Transition to a Conducting State.
\newblock \emph{Phys. Rev. Lett.} {\bf 116}, 255501 (2016).

\bibitem{Plorenzen}
\bibinfo{author}{Lorenzen, W.}, \bibinfo{author}{Holst, B.} \&
  \bibinfo{author}{Redmer, R.}
\newblock \bibinfo{title}{First-order liquid-liquid phase transition in dense
  hydrogen}.
\newblock \emph{\bibinfo{journal}{Phys. Rev. B}} \textbf{\bibinfo{volume}{82}},
  \bibinfo{pages}{195107} (\bibinfo{year}{2010}).

\bibitem{HM2016}
\bibinfo{author}{{Hubbard}, W.~B.} \& \bibinfo{author}{{Militzer}, B.}
\newblock \bibinfo{title}{{A Preliminary Jupiter Model}}.
\newblock \emph{\bibinfo{journal}{ApJ}} \textbf{\bibinfo{volume}{820}},
  \bibinfo{pages}{80} (\bibinfo{year}{2016}).
\newblock \eprint{1602.05143}.

\bibitem{HG2013}
\bibinfo{author}{{Helled}, R.} \& \bibinfo{author}{{Guillot}, T.}
\newblock \bibinfo{title}{{Interior Models of Saturn: Including the
  Uncertainties in Shape and Rotation}}.
\newblock \emph{\bibinfo{journal}{The Astrophysical Journal}}
  \textbf{\bibinfo{volume}{767}}, \bibinfo{pages}{113} (\bibinfo{year}{2013}).
\newblock \eprint{1302.6690}.

\bibitem{Motta_benchmark}
\bibinfo{author}{Motta, M.} \emph{et~al.}
\newblock \bibinfo{title}{{Towards the solution of the many-electron problem in
  real materials: equation of state of the hydrogen chain with state-of-the-art
  many-body methods}}.
\newblock \emph{\bibinfo{journal}{A\&A}} \textbf{\bibinfo{volume}{596}},
  \bibinfo{pages}{A114} (\bibinfo{year}{2017}).
\newblock \eprint{1705.01608}.

\bibitem{refkturbo}
\bibinfo{author}{Dagrada, M.}, \bibinfo{author}{Karakuzu, S.},
  \bibinfo{author}{Vildosola, V.~L.}, \bibinfo{author}{Casula, M.} \&
  \bibinfo{author}{Sorella, S.}
\newblock \bibinfo{title}{Exact special twist method for quantum monte carlo
  simulations}.
\newblock \emph{\bibinfo{journal}{Phys. Rev. B}} \textbf{\bibinfo{volume}{94}},
  \bibinfo{pages}{245108} (\bibinfo{year}{2016}).

\bibitem{ccpVXZ}
\bibinfo{author}{Jr., T. H.~D.}
\newblock \bibinfo{title}{Gaussian basis sets for use in correlated molecular
  calculations. i. the atoms boron through neon and hydrogen}.
\newblock \emph{\bibinfo{journal}{The Journal of chemical physics}}
  \textbf{\bibinfo{volume}{90}}, \bibinfo{pages}{1007} (\bibinfo{year}{1989}).

\bibitem{QE-2009}
\bibinfo{author}{Giannozzi, P.} \emph{et~al.}
\newblock \bibinfo{title}{Quantum espresso: a modular and open-source software
  project for quantum simulations of materials}.
\newblock \emph{\bibinfo{journal}{Journal of Physics: Condensed Matter}}
  \textbf{\bibinfo{volume}{21}}, \bibinfo{pages}{395502 (19pp)}
  (\bibinfo{year}{2009}).
\newblock \urlprefix\url{http://www.quantum-espresso.org}.

\end{thebibliography}
\end{document}